\documentclass[fleqn,10pt]{wlscirep}
\usepackage[utf8]{inputenc}
\usepackage[T1]{fontenc}
\usepackage{siunitx}
\usepackage{gensymb}
\usepackage{amsmath}
\usepackage{graphicx}
\usepackage{wrapfig}
\usepackage{lipsum}
\usepackage[font={sf,small}]{caption}

\title{Rapid Design and Fabrication of Body Conformable Surfaces with Kirigami Cutting and Machine Learning}

\author[1+]{Jyotshna Bali}
\author[1+]{Jinyang Li}
\author[1, 2, 3]{Jie Chen}
\author[1, 3*]{Suyi Li}
\affil[1]{Department of Mechanical Engineering, Virginia Tech, Blacksburg, VA 24061, USA}
\affil[2]{Macromolecules Innovation Institute, Virginia Tech, Blacksburg, VA 24061, USA}
\affil[3]{VT Made, Virginia Tech, Blacksburg, VA 24061, USA}
\affil[*]{Corresponding Authors: \texttt{jyotshna@vt.edu, suyili@vt.edu}}
\affil[+]{These authors contributed equally to this work}

\begin{abstract}

By integrating the principles of kirigami cutting and data-driven modeling, this study aims to develop a personalized, rapid, and low-cost design and fabrication pipeline for creating body-conformable surfaces around the knee joint. The process begins with 3D scanning of the anterior knee surface of human subjects, followed by extracting the corresponding skin deformation between two joint angles in terms of longitudinal strain and Poisson’s ratio. In parallel, a machine learning model is constructed using extensive simulation data from experimentally calibrated finite element analysis. This model employs Gaussian Process (GP) regression to relate kirigami cut lengths to the resulting longitudinal strain and Poisson’s ratio. With an R$^2$ score of 0.996, GP regression outperforms other models in predicting kirigami’s large deformations. Finally, an inverse design approach based on the Covariance Matrix Adaptation Evolution Strategy (CMA-ES) is used to generate kirigami patch designs that replicate the in-plane skin deformation observed from the knee scans. This pipeline was applied to three human subjects, and the resulting kirigami knee patches were fabricated using rapid laser cutting, requiring only a business day from knee scanning to kirigami patch delivery. The low-cost, personalized kirigami patches successfully conformed to over 75\% of the skin area across all subjects, establishing a foundation for a wide range of wearable devices. The study demonstrates this potential through an impact-resistant kirigami foam patch, which not only conforms to dynamic knee motion but also provides joint protection against impact. Finally, the proposed design and fabrication framework is generalizable and can be extended to other deforming body surfaces, enabling the creation of personalized wearables such as protective gear, breathable adhesives, and body-conformable electronics.

\medskip
\textbf{keywords}: Kirigami, Wearables, Surrogate Modeling

\end{abstract}
\begin{document}

\flushbottom
\maketitle

\thispagestyle{empty}

\section{Introduction}

Over the past decades, body-conformable and wearable devices have attracted wide interest from many communities.  These wearable devices aim to mimic the mechanical properties of the underlying skin, enabling them to conform to its non-uniform deformation and irregular shape. As a result, wearable devices can operate with minimal interference with the dynamic body motions, while ensuring bio-compatibility, functional performance, comfort, and durability. 
A large subset of the conformable devices takes the form of a surface with different thicknesses. So matching their in-place surface deformations with the underlying skin plays a critical role. To this end, researchers have developed two complementary approaches --- structural design and material engineering \cite{shimura2023engineering, xu2020skin}. 
The structural design approach employs tailored geometric transformation via structure patterning techniques, such as island bridges and network architectures \cite{lim2024material} using filamentary serpentine and wavy ribbons \cite{kim2019soft, kim2011epidermal, kim2014stretchable}, fractal-inspired and hierarchical motifs \cite{fan2014fractal, xu2013stretchable}, spirals and springs \cite{sung2015development, rojas2014design},  helical coils \cite{yan2023hierarchical}, honeycomb grids and chains \cite{lee2018nonthrombogenic, li2020wearable}, auxetics \cite{liu2025efficient}, Origami \cite{hou2024programmable, zhou2025hyper}, and Kirigami \cite{tao2023engineering, chow2024soft, jin2024engineering}. 
The material engineering approach, on the other hand, exploits the constituent material's stretchability and adhesion \cite{choi2025motion, chen2024customized} to improve conformability. This approach employs elastomeric polymers and composites \cite{guan2020air, lv2020flexible, zhang2020fully}, hydrogel polymers \cite{lim2021tissue}, liquid metals \cite{ma2023shaping, li2023ultra}, shape memory polymers \cite{chitrakar2023multifaceted}, and fabrics \cite{xiang2024recent, phan2022smart}.
These structural design and material engineering techniques can be seamlessly combined with advanced manufacturing and integration technologies \cite{zhang2023recent, wei2024revolutionizing}, such as printing and coating \cite{park2022high, raghav2023fabrication}, thin film deposition and transfer printing \cite{gao2019flexible}, as well as self-assembly and morphing \cite{mirzababaei20243d}. 
Implementing these skin-conformable surfaces can advance healthcare and biomedical wearables \cite{lin2024stretchable, deng2023smart, lim2024material, zhao20243d}, soft robotics and human-machine interfaces \cite{xiong2021functional}, energy harvesting and wearable power systems \cite{hou2024programmable, saifi2024ultraflexible, zhang2022human}, smart textiles and fashion technologies \cite{wicaksono2020tailored, phan2022smart, zhao20243d}, as well as sports and performance monitoring systems \cite{sun2022review, zhao20243d}.

While surface conformability in wearable devices have significantly enhanced their functionalities, several challenges still limit their widespread applications. Mechanical mismatch is one of the most prominent ones, and it often results in discomfort to the users and significantly reduces the wearable performance \cite{shimura2023engineering}. This challenge highlights the importance of incorporating \textit{personalized} and \textit{accurate} skin deformation knowledge into the wearable design \cite{xu2020skin}. 
Few state-of-the-art wearable devices are personalized --- most of them adopted a non-ideal  ``one size fits all'' strategy, assuming that in-plane softness is sufficient for conformability. (That is, one typically assumes that ``if the wearables are sufficiently soft, they should always fit the human body.'') Such simplification inevitably leads to performance tradeoffs and mechanical mismatches. On the other hand, designing for individual human variability is an enormous task with a seemingly infinite variable space and complex objectives. As a result, personalized design can be quite time-consuming and resource-intensive.  

Therefore, in this study, we aim to formulate a design and fabrication pipeline to overcome these challenges and create low-cost, highly individualized, and body-conformable wearable surfaces.  To this end, we will adopt Kirigami cutting and data-driven surrogate modeling techniques and focus on the knee surface as the case study. 
The principle of Kirigami cutting has been widely used to impart flexibility, stretchability, conformal contact, adhesion, and programmable shape change to over-the-body applications, such as wearable bio-electrodes, electronic skins, sensors \cite{won2019stretchable, chow2024soft}, implantable \cite{fernandez2021body, lee2023natural}, e-textile \cite{li2019Kirigami}, thermo-regulation devices \cite{chen2023Kirigami}, as well as other multifunctional integrated platforms \cite{tao2023engineering}. Kirigami-cut surfaces have a unique ability to stretch anisotropically through the opening of their cuts and internal deformation, which can be designed to match the local skin deformation underneath while maintaining contact and breathability. This ability has been utilized to design wearable patches around movable joints, such as the shoulder \cite{alkayyali2024identifying}, knee {\cite{zhao2018Kirigami, yang2024knee}}, elbow {\cite{wang2021design}}, and wrist {\cite{kang2025skin, abbasipour2023stretchable}}. However, these Kirigami patches are either manufactured with minimal individualization or lack the rapid design and fabrication potential for scaling up. 

\begin{figure}
    \centering
    \includegraphics[width=\textwidth]{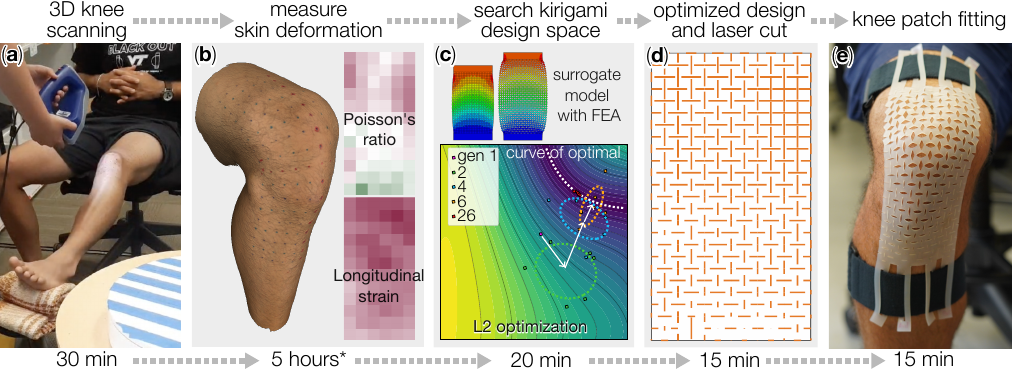}
    \caption{Flowchart of the rapid design and fabrication pipeline of personalized Kirigami knee patch.  Note that the second step --- measure skin deformation --- is completed manually in this study due to the need to manually transfer data between different software. Developing an integrated and automated software package could further speed up this process.}
    \label{fig: flowchart}
\end{figure}

In this study, we achieved the rapid design and fabrication of conformable knee patches by carefully tuning their cutting parameters (i.e., the length of every cut) to accurately match the complex 3D skin deformation. 
More specifically, we first gathered the skin deformation data by 3D scanning the subject's knee surface at the initial and final knee joint angles (Figure~\ref{fig: flowchart}a). These scanning data allowed us to quickly extract skin deformation information, including longitudinal strain and Poisson's ratio (heatmaps in Figure~\ref{fig: flowchart}b).
Meanwhile, we conducted Finite Element Analysis (FEA) and physical testing of silicon rubber Kirigami sheets with a uniform cut pattern to correlate the cut parameters to their elastic deformation. These experimentally-validated FEA data enabled us to construct an accurate, high-resolution surrogate model for inverse design (Figure~\ref{fig: flowchart}c). We compared multiple machine learning models, Gaussian Process regression (GP, also known as Krigin) \cite{marrel2024probabilistic, sun2024many, wu2024risk}, Polynomial regression (PR), Gradient Boosting Regression (GBR), and Random Forest Regression (RF), among which GP is the most accurate for the Kirigami design problem. 
Finally, we adopted a state-of-the-art optimization algorithm, Covariance Matrix Adaptation Evolution Strategy (CMA-ES) \cite{li2023elite} to search for the optimal Kirigami cut designs that conform to skin deformation. The optimized design can be quickly manufactured using a laser cutter (Figure~\ref{fig: flowchart}d, e). 

Thanks to the design flexibility and simple fabrication of Kirigami, the design and fabrication pipeline can be i) easily personalized because only a simple 3D scanning is required, ii) with a low cost because laser cutting is widely accessible, and iii) rapid so that it takes only a working day (about 6.5 hours) between knee scanning and the Kirigami patch completion. 

The rest of the paper details the different steps of the Kirigami knee patch design and fabrication pipeline, discusses the outcome, and outlines its future applicability. 

%\newpage

\section{Methodology}

\subsection{Obtaining The Design Target: User's Knee Skin Deformation}
%\vspace{5mm}

\begin{figure}[t!]
    \centering
    %\vspace{-20pt}
    \includegraphics[scale=1.0]{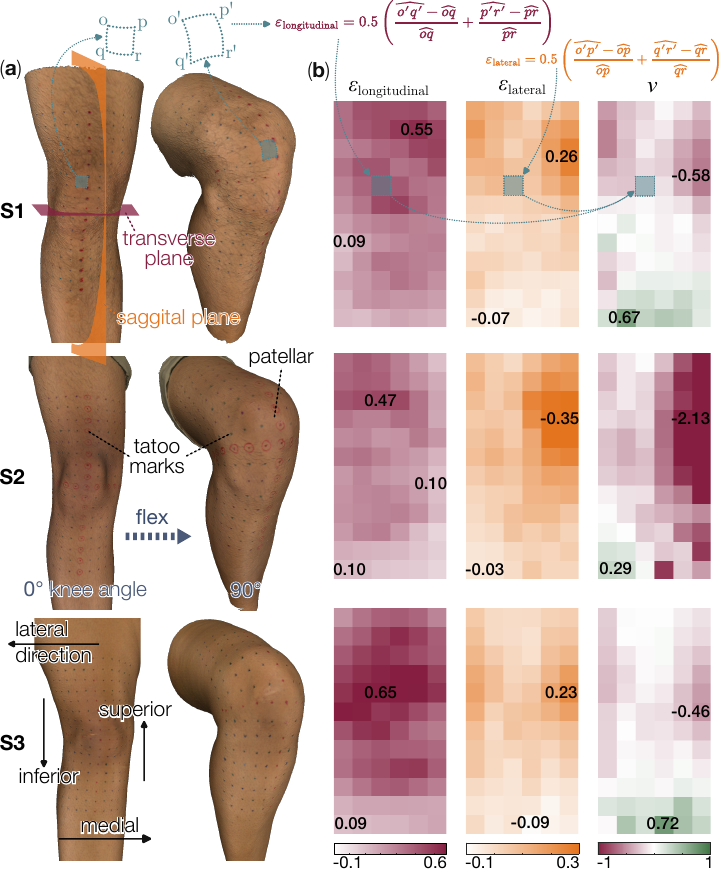}
    \caption{Scanning and measuring the subjects' knee surface deformation. (a) Reconstructed knee surfaces of the 3 subjects (i.e., S1, S2, and S3), using the 3D scanning data. (b) Heatmaps showing the longitudinal, lateral strain, and Poisson's ratio distribution from the 0\degree~to 90\degree~knee flex. The unit cells with the maximum and minimum values in each heatmap are labeled, further highlighting the variety between different subjects. Note that the length variables (e.g., $\widehat{op}$ and $\widehat{q'r'}$) in the strain calculations are geodesic distances on the 3D knee surface.}
    \label{fig: scan}
\end{figure}

To conduct this human participant research, we first obtained approval from the Virginia Tech Institutional Review Board (IRB No. 24-775) and written consent from all the volunteers. Then, to measure the skin deformation in an accurate, reliable, and consistent setup, we used a temporary tattoo marker to draw a grid directly on three volunteer subjects' right knee skin (Figure~\ref{fig: scan}a). The grid has a size of $240\times180$mm with a square unit cell size of 15mm, totaling 16 rows and 12 columns. It spans from the medial to the lateral epicondyles and extends uniformly from the superior to the inferior region of the patella. 
The skin deformation pattern was then captured with a high-resolution, handheld 3D Scanner (Artec Eva), as the subjects flexed their knee joints from a relaxed \ang{0} to the \ang{90} angle. 
The scanner generated a dense point cloud, which was then converted into a smoothed polygon model and exported in the OBJ format. Then, we post-processed the data in the open-source GigaMesh software to measure the geodesic distances between adjacent marker points~\cite{VAST:VAST10:131-138}.  By comparing these geodesic distances before and after knee flexure, we can calculate the averaged longitudinal strain, lateral strain, and Poisson's ratio of each cell. 
Overall, this process allowed us to map the 3D surface scans of the knee skin into a flattened 2D strain and Poisson's ratio heatmap (Figure \ref{fig: scan}b). 

The longitudinal strain heatmap of all three subjects showed positive stretches in almost all regions of the knee, which agrees with the established literature on skin deformation of the lower body {\cite{wessendorf2012dynamic, choi20153d}}. Such longitudinal stretch is particularly prominent around the patellar region, as indicated by a darker maroon color in the first column of heatmaps in Figure \ref{fig: scan}(b). 
In comparison, the magnitude of lateral strain appeared roughly halved, and a greater lateral stretch, as denoted by the orange color, was observed on both medial and lateral sides of the patella in all three subjects. Notably, the regions below the patella exhibited small lateral compression, which concentrated on the most inferior portion of the scanned region.  
Correspondingly, the Poisson's ratio heatmap shows predominantly negative values (third column of heatmaps in Figure \ref{fig: scan}b), indicating that the knee skins are expanding in their surface area during knee joint flexing. Meanwhile, a small region of positive Poisson's ratio --- corresponding to lateral compression and longitudinal expansion --- occurs along the inferior portion of the knee scan (green regions in this heatmap). 
In what follows, the longitudinal strain and Poisson's ratio heatmaps will serve as the objective functions for the Kirigami design.

\subsection{Constructing The Kirigami's Design Space}

\begin{figure}[h!]
    \centering
    %\vspace{-100pt}
    \includegraphics[scale=0.87]{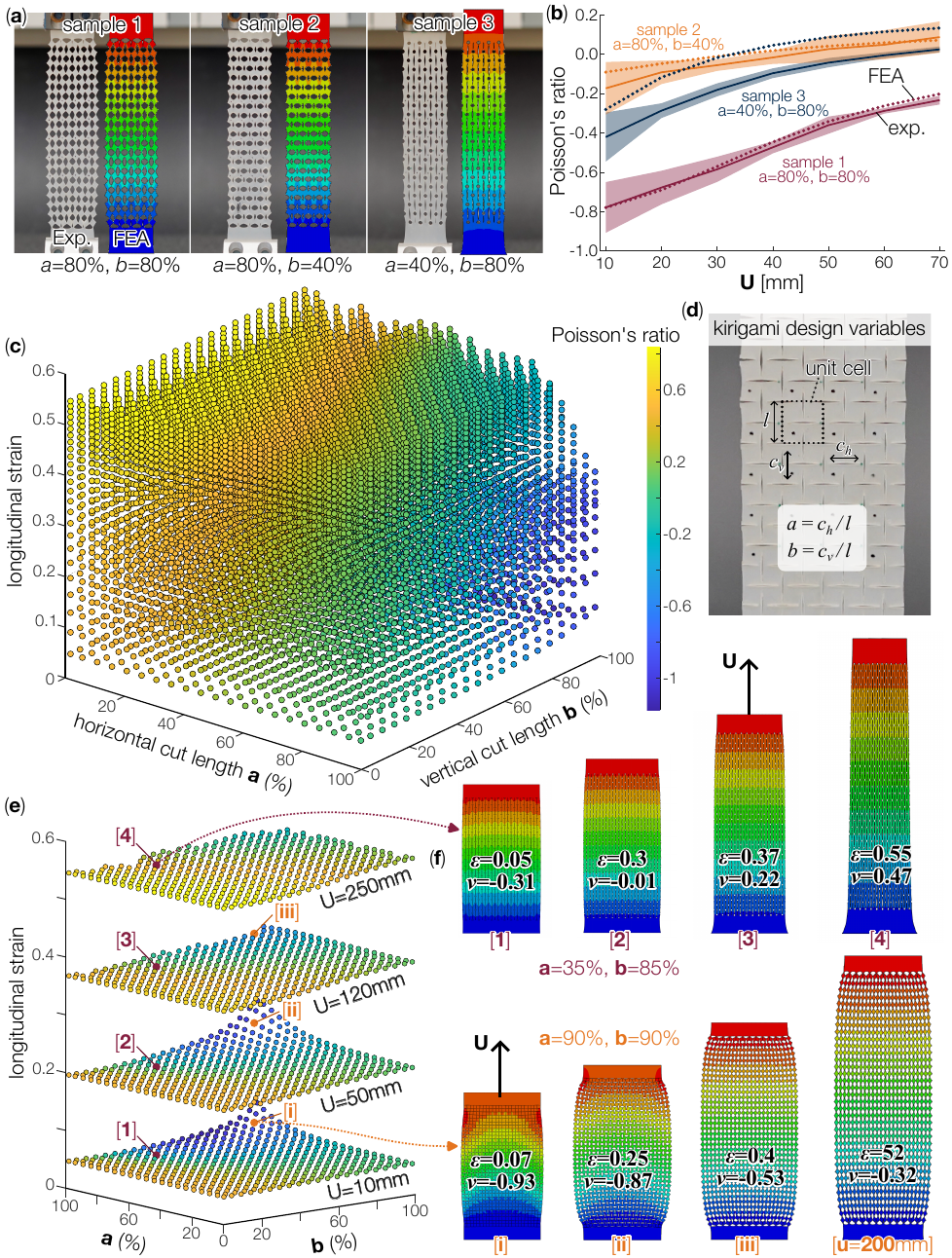}
    \caption{Populating the design space of Kirigami with experimentally-calibrated finite element analysis (FEA). (a) Comparing the Kirigami sheets' deformation pattern between FEA and experiment (with top end displacement U=70mm). (b) Comparing the Poisson's ratio between FEA and experiment. Here, the solid lines are averaged test data, and the shaded bands are standard deviations from the readings of 12 unit cells in the middle of the samples. The dashed lines are FEA predictions. (c) Populated design space, where each marker corresponds to a unique combination of Kirigami cut design and longitudinal strain, and the color of the marker represents the corresponding Poisson's ratio. (d) A close-up view of the Kirigami design variables. (e) A subsection of this design space, showing a few slices corresponding to U=10, 50, 120, 250mm. (f) Detailed FEA simulation outcomes, where [i-iv] and [1-4] highlight two Kirigami designs.}
    \label{fig: FEA}
\end{figure}

The skin deformation heatmaps from the three subjects showed a wide range of longitudinal strain and Poisson's ratio combinations. Therefore, in this section, we explore whether the Kirigami sheet can capture such variation with intentional design.  To this end, we adopted the classical Kirigami design with a tessellated ``cross cut'' pattern (Figure~\ref{fig: FEA}).  This pattern can be divided into square-shaped unit cells, which conveniently correspond to the grid pattern used in the 3D knee surface scanning.  For each unit cell, the lengths of its horizontal and vertical cuts are denoted as $a$ and $b$, respectively. They are percentage values indicating the cut length with respect to the unit cell size (Figure~\ref{fig: FEA}d).

For example, the three Kirigami samples shown in Figure~\ref{fig: FEA}(a) were made from a $140\times40\times2$ mm silicon rubber sheet (Dragon Skin\texttrademark 20), consisting of an array of square unit cells with $10\times10$ mm in size. The first sample had a uniform cut pattern of $a=80\%$ and $b=80\%$, meaning the lengths of horizontal and vertical cuts are both 80\% of the unit cell size. In contrast, the second and third samples had unequal horizontal and vertical cut lengths. We then clamped the Kirigami sheets at both ends and uniformly stretched them on a universal tensile tester machine (Instron 6800 Series with a 100N load cell). To measure the longitudinal strain and Poisson's ratio with minimal boundary effects, we focused on the cells in the middle and captured their deformed shapes with a high-resolution camera (Sony $\alpha$7C II). These images can be analyzed with the image processing toolbox in MATLAB to obtain the deformation curves shown in Figure~\ref{fig: FEA}(b). 

In parallel to the experiments, we also conducted finite element analysis (FEA) of the same Kirigami samples using 2D shell elements (ABAQUS, CPS4R) and Neo-Hookean constitutive properties (shear modulus, $ \mu = 0.207 MPa$ \cite{marechal2021toward}).  In these FEA models, the Kirigami cuts were represented by the ``seam'' feature, which partitions the sheet horizontally and vertically to allow structured meshing (seed size 0.5).  Similar to the experiment setup, we applied longitudinal stretching displacement at the top boundary, while fully fixing the bottom boundary.  The longitudinal strain and Poisson's ratio of the unit cells in the middle region of this simulated Kirigami sheet were then compared with the experimental data, which showed a good agreement (Figure~\ref{fig: FEA}b).  Therefore, the finite element model can reliably capture the Kirigami sheet's deformation characteristics, so we could use it to investigate different Kirigami cross-cut patterns and explore the full design space. 

To this end, we conducted extensive simulations to populate Kirigami's design space. We modeled and simulated 441 Kirigami sheets that have the same overall size but different cut lengths (Figure~\ref{fig: FEA}c-e).  These Kirigami sheets were $220\times120$mm, consisting of 18 rows and 12 columns of $10\times10$mm unit cells. Within each Kirigami sheet, the cut lengths are uniform, but between the different Kirigami sheets, the horizontal and vertical cut lengths vary independently from $5 \%$ to $95 \%$ of the unit cell size, with an increment of $5\%$.  In addition, $1\%$ and $99\%$ cut lengths were included to define the minimum and maximum cut limits of the design space. We then simulated these Kirigami sheets' longitudinal deformation under 10 to 250mm of stretching displacements, with increments of 10mm. In every simulation, the deformations of the middle four unit cells were used to calculate the output longitudinal strain and Poisson's ratio. Therefore, the output of these 441 simulations populate the full design space of Kirigami, as shown in Figure~\ref{fig: FEA}(c).

This full design space plot confirmed that, with a tailored horizontal and vertical cut length and appropriate stretching, the Kirigami sheet can achieve a wide range of longitudinal strain and Poisson's ratio combinations. Here, each circular marker represents a Kirigami sheet of a particular horizontal and vertical cut length, pulled by prescribed stretching displacement. As a Kirigami sheet is being stretched, its longitudinal strain increases at a greater rate at first, and then slows down later after most of the cuts have opened up. Interestingly, Poisson's ratio increases monotonically as its longitudinal strain increases, which is reflected by the color distribution of the circular markers. Kirigami sheets of smaller horizontal and vertical cuts can exhibit a positive and increasing Poisson's ratio throughout their deformation range. On the other hand, Kirigami sheets with longer cut lengths could show a transformation from being auxetic to non-auxetic as their stretch increases. Such a trend can be more visible by slicing and highlighting several layers of the 3D Design space (e.g., 10, 50, 120, and 250 mm deformation level as shown in the Figure~\ref{fig: FEA}(e). 

To better illustrate the Kirigami's deformation characteristics, Figure~\ref{fig: FEA}(f) details the simulation outcomes of two samples  --- one has a moderate cut length $a=35\%, b=85\%$, and the other has a longer cut at $a=90\%,b=90\%$.  As the former Kirigami is stretched from 0.05 to 0.55 strain levels, its rectangular ``facets'' would rotate, thus expanding the lateral dimensions and creating the classical auxetic behavior.  In other words, the kinematics of facet rotation dominate the deformation behavior at low stretching strain. However, as this Kirigami's stretch increases beyond the 0.55 strain level, the facet rotation reaches its kinematic limits. As a result, the facets stop rotating, but become elastically deformed, exhibiting the positive Poisson's ratio. Importantly, longer Kirigami cuts increase the kinematic limit of facet rotation.  For example, the other Kirigami sheet with a 90\% cut is stretched significantly by 200mm in our simulation. Its Poisson's ratio increases from -0.93 to -0.32; however, it did not reach the kinematic facet rotation limit. 

In summary, with the help of extensive FEA simulations of Kirigami sheets with varying horizontal and vertical cut lengths, we were able to fully populate their design space.  This space will serve as the foundation for conformable surface design in the following sections. 

\subsection{Surrogate Modeling of The Kirigami Structure–Property Relationship}
While the populated design space from the finite element model was comprehensive, it was still sampled by discrete cut lengths with 5\% increments. Such a non-continuous nature can present limitations for inverse designing the Kirigami cut based on knee skin deformation. For example, precisely matching the knee skin deformation might require cut lengths not sampled by the finite element simulations. Moreover, multiple Kirigami cut lengths could give similar deformation characteristics to match the knee skin deformation (i.e., many-to-one mapping). One can certainly address this issue by running more FEA simulations to increase the resolution of the design space, but the computational cost can be prohibitive. Therefore, we built a surrogate model using machine learning methods to accurately search Kirigami cut sizes with the least deviation from the design objectives.

\begin{table}[b]
    \centering
    \caption{Performance comparison of surrogate models}
    \label{tab: Accuracies of candidate surrogate models}
    \begin{tabular}{l c}
        \toprule
        \textbf{surrogate model} & \textbf{R\textsuperscript{2}} \\
        \midrule
        Gaussian Process Regression       & 0.9969 \\
        Gradient Boosting Regression      & 0.7961 \\
        Random Forest Regression          & 0.5120 \\
        Polynomial Regression (degree=3)  & 0.9124 \\
        \bottomrule
    \end{tabular}
\end{table}

\begin{figure}[ht!]
    \centering
    \includegraphics[scale=0.9]{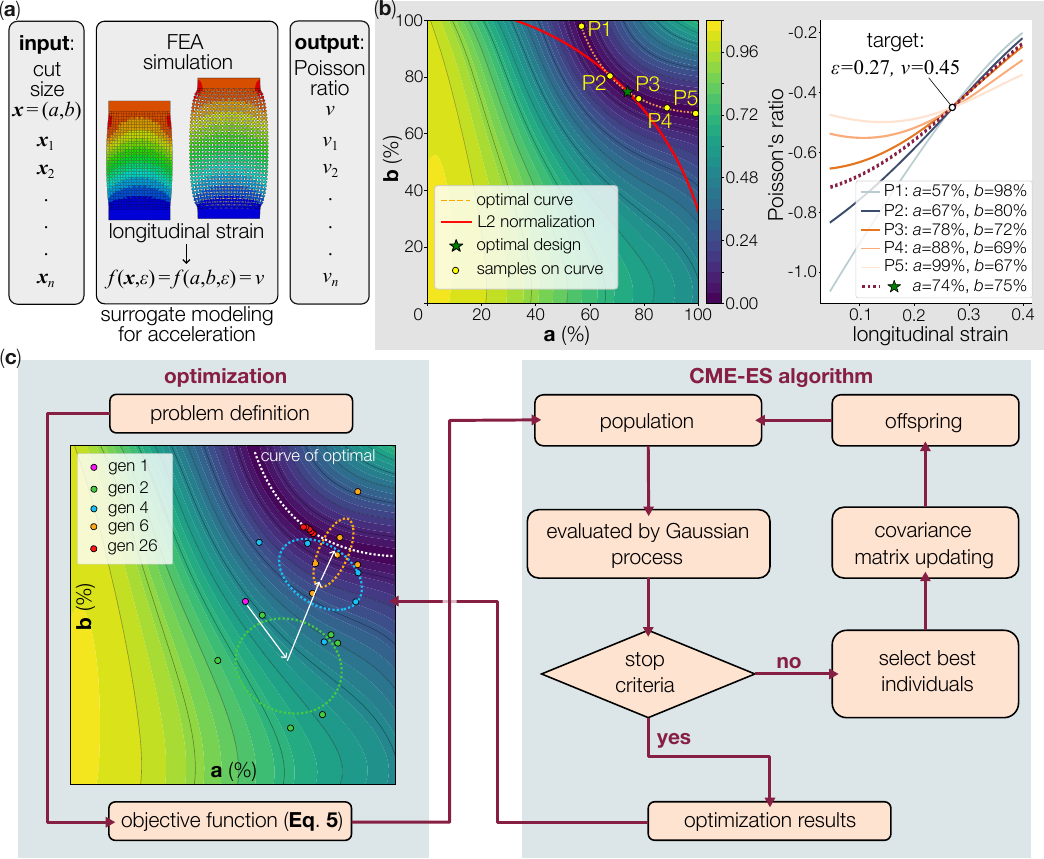}
    \caption{Elements of the inverse design methodology. (a) The input and output architecture of the surrogate modeling. (b) L2-Normalization for design candidate selection. In the design space plot on the left, we selected 6 design candidates on the optimal curve (i.e., P1-P5 and the optimal candidate). Then their longitudinal strain vs. Poisson's ratio responses are plotted on the right. (c) The flow chart of the optimal Kirigami design: The initial data was sampled randomly at first, as shown by generation 1. After a few iterations, it converges to our optimal solution at the 26th iteration.}
    \label{fig: optimization}
\end{figure}

\textbf{Variables shifting for surrogate modeling:} We attempted to use the finite element simulation outcome to generate the surrogate model. However, there is no one-to-one mapping from the geometric design variables, i.e., $a$ and $b$, to the targeted deformation characteristics, i.e., longitudinal strain and Poisson's ratios. Considering the longitudinal strain increases monotonously in the tensile simulation, we decided to use it as another input to the surrogate model in addition to the two cut length variables. As a result, the surrogate model will output the Poisson's Ratio as the prediction (Figure~\ref{fig: optimization}a).

To ensure accuracy, we compared the following regression models: Gaussian Process Regression (GP, also known as Kriging), Random Forest Regression (RF), Polynomial Regression (PR), and Gradient Boosting Regression (GBR). As shown in Table~\ref{tab: Accuracies of candidate surrogate models}, 
the GP provides the highest accuracy. Our surrogate modeling can be represented as follows:
\begin{equation}
    \mu = FEM(a_x,b_y,\epsilon)\approx f(a_x,b_y,\epsilon),
\end{equation} 
where $f$ represents the surrogate models, we use the cross-validation method to test its accuracy in 10 folds using the 441 samples. In 10-fold cross-validation, the dataset is split into 10 equal parts. Then, we process the modeling 10 times, with each part being the test set once. Finally, the modeling accuracy is evaluated as the mean of the 10 models. The model accuracy is shown in Table~\ref {tab: Accuracies of candidate surrogate models}. Clearly, the Gaussian Process Regression presents the most accurate prediction ($R^2=0.997$), so we use it as the basis of the surrogate model hereafter. For readability, we provide a brief introduction of GP. Additional details are available in Ref. \cite{seeger2004gaussian}.
\begin{equation}
\mu(\mathbf{x}^*) =GP(x^*)=GP(a_x,b_y,\epsilon)= \mathbf{k}^\top \left( \mathbf{K} + \sigma_n^2 \mathbf{I} \right)^{-1} \mathbf{y},
\end{equation}
where $\mu(x^*)$ is the mean value of the response of the input data point $x^*$. In Equation (2), $\mathbf{k}^\top$ and $\mathbf{K}$ are calculated from the Gaussian kernel function. $\mathbf{K}$ is a matrix, and its elements can be represented by
\begin{equation}
    K_{ij}=k(x_i,x_j)=\sigma_n^2\exp(-\frac{\|x_i-x_j\|^2_2}{2l^2}),
\end{equation} 
with some hyper parameters $\sigma,l$ where $x_i,x_j$ are the input data sampled from the training dataset and $\mathbf{y}=\begin{bmatrix}
y_1 & y_2 & \dots &y_n \end{bmatrix}^\top$ are the responses to the sampled data. For $\mathbf{k}^\top$, it calculates the distance to all sampled data points,
\begin{equation}
    \mathbf{k}^\top=\begin{bmatrix}
        k(x^*,x_1)&k(x^*,x_2) & \dots & k(x^*,x_n))
    \end{bmatrix}^\top.
\end{equation}

\subsection{Inverse Design And Fabrication of The Kirigami Knee Patches}
%\vspace{2mm}

With the accurate surrogate model, we can perform inverse design using optimization techniques. Although we simulated extreme values of design variables (i.e., $a=99\%$, and $b=99\%$), it is nearly impractical to fabricate them. As a result, the maximum horizontal and vertical cut lengths are limited to 90\%. More importantly, our surrogate model reveals that, for a given combination of longitudinal strain $\varepsilon$ and Poisson's ratio $\nu$, there are an infinite number of cut length solutions. To identify a unique and optimal design, we introduce the L2-normalization --- a technique in machine learning to overcome overfitting. For example, Figure~\ref{fig: optimization}(b) shows a curve of optimal solutions, where all the Kirigami cut designs on this curve can minimize our objective function: $\varepsilon=0.27, \nu=0.45$. However, some of these points are at the boundary of the design space with a large horizontal cut and a small vertical cut, or vice versa. This may cause excessive stress concentration. To solve this problem, L2-normalization selects the solution that is closest to the origin of the design space to maintain relatively equal cut lengths in both directions. In Figure~\ref{fig: optimization}(b), we selected 6 design candidates on the optimal curve, including the L2-normalized design, and plot their longitudinal strain vs. Poisson's ratio responses. All these designs can satisfy our design objective; however, the L2-normalization selects the design closest to the origin, giving the most even cut lengths.

Designing all targets consecutively requires about 4 hours. However, with parallel computation, the process is completed in only 20 minutes.

\begin{figure}[t]
    \centering
    \includegraphics[scale=0.9]{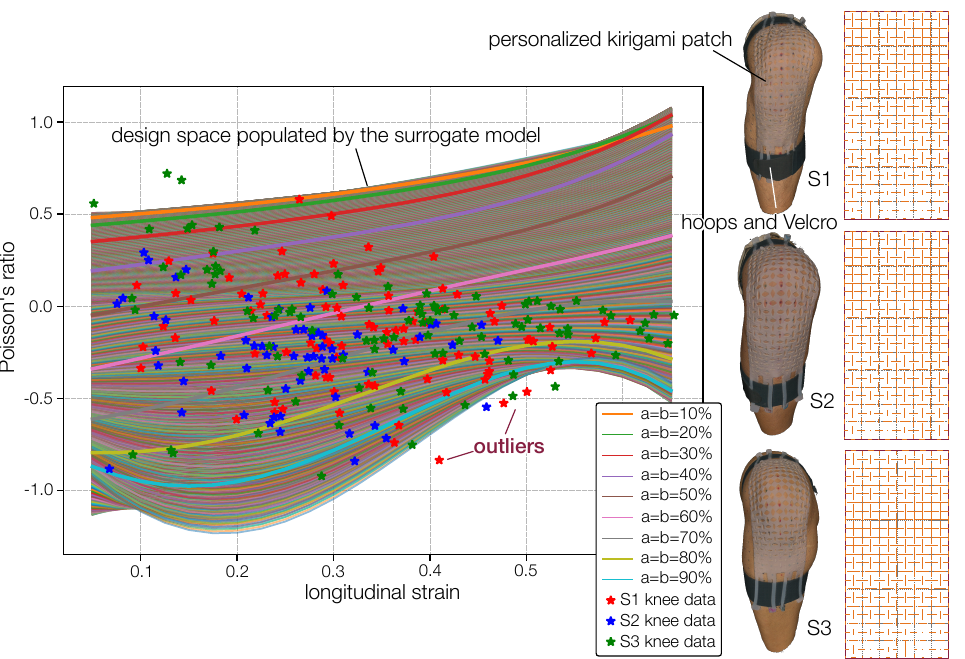}
    \caption{The final, personalized Kirigami patches for the three subjects. (left) In this plot, each star-shaped marker represents the longitudinal strain--Poisson's ratio combination of a grid cell on the subject's knee surface. They are the same data from the heatmaps in Figure~\ref{fig: scan}(b), but plotted differently. Meanwhile, each thin curve in this plot is the longitudinal strain--Poisson's ratio response corresponding to a unique Kirigami cut design. These curves are generated by the surrogate model, which is based on the FEA simulation data in Figure~\ref{fig: FEA}(c). Essentially, the optimization procedure laid out in Figure~\ref{fig: optimization}(c) finds the curve that best matches each star-shaped marker. (right) The final Kirigami patch designs. The images in the left column are the reconstructed 3D scanning data from the subjects wearing their Kirigami patches, and the images in the right column are the corresponding Kirigami designs.}
    \label{fig: output}
\end{figure}

After establishing the surrogate model and design optimization procedures, we proceed to inverse design of the Kirigami cut according to the 3D scanned skin deformation. That is, the 3D knee scanning data produced a set of heat maps containing $12\times6$ (row x column) grid cells, each has a unique combination of longitudinal strain and Poisson's ratio (Figure~\ref{fig: scan}). Correspondingly, the Kirigami knee patch should also have $12\times6$ unit cells. The cut lengths of each cell should be selected to match the corresponding knee scan data using the aforementioned surrogate model and design optimization procedures (Figures~\ref{fig: optimization}).

To this end, we first compared the target unit cell deformations from the knee scan data and the achievable unit cell deformation from the Kirigami surrogate model.  By plotting these two sets of data together (Figure~\ref{fig: output}, left), we found that the large design space of Kirigami could indeed cover the variability of subjects' skin deformations. While outliers exist (aka, there are several unit cells on subjects' knee showing deformations beyond Kirigami's design space), they do not venture far away from Kirigami's capability. 
Therefore, we could formulate the personalized Kirigami design optimization as:
\begin{equation}
\begin{aligned}
    \min_{a,b}:\quad & l = \left(\mu_{\text{scan}} - GP(a,b,\epsilon_{\text{scan}})\right)^2 + \lambda(a^2 + b^2) \\
    \text{s.t.:}\quad & 0 \leq a \leq 90\%,\quad 0 \leq b \leq 90\% 
\end{aligned}
\end{equation}
where $l$ is the objective function measuring the difference between the targeted Poisson's ratio from the 3D knee scan data and the predicted Poisson's ratio from the surrogate model, and $\lambda$ is the penalty coefficient of the L2 normalization. The above optimization process is used to design the cut lengths of all unit cells in the Kirigami patch. The optimization problem was solved using the state-of-the-art optimization CMA-ES, which encodes the input variables and forms a population to initiate the search process. After selection, crossover, and mutation operators to generate the offspring population to optimize the objective function, the process terminates upon reaching the specified stopping criteria, yielding the current optimal solution (Figure~\ref{fig: optimization}c).

Since the knee surface's longitudinal strain and Poisson's ratio can differ significantly between adjacent grid cells, the corresponding Kirigami cut length also differs between adjacent unit cells.  Therefore, for the cuts at the boundary between adjacent cells, we performed a simple averaging and skewing.  The final design was exported as a vector image file for laser cutting. To fabricate the Kirigami knee patch, we first prepared a 2mm thin silicon rubber sheet with the help of a film applicator (DragonSkin 20). After curing, the rubber sheet is cut with a CO$_2$ laser cutter (Trotec Speedy 360, 40\% of max power, 0.4\% of max laser head speed, and 1000~Hz excitation frequency). Note that several additional hoops were designed at the two ends of the Kirigami patch so that we can use Velcro straps to quickly attach the Kirigami patch to the subjects' knees (Figure~\ref{fig: output}, right).

\section{Results and Discussion}

\begin{figure}[t!]
    \centering
    \includegraphics[scale=0.9]{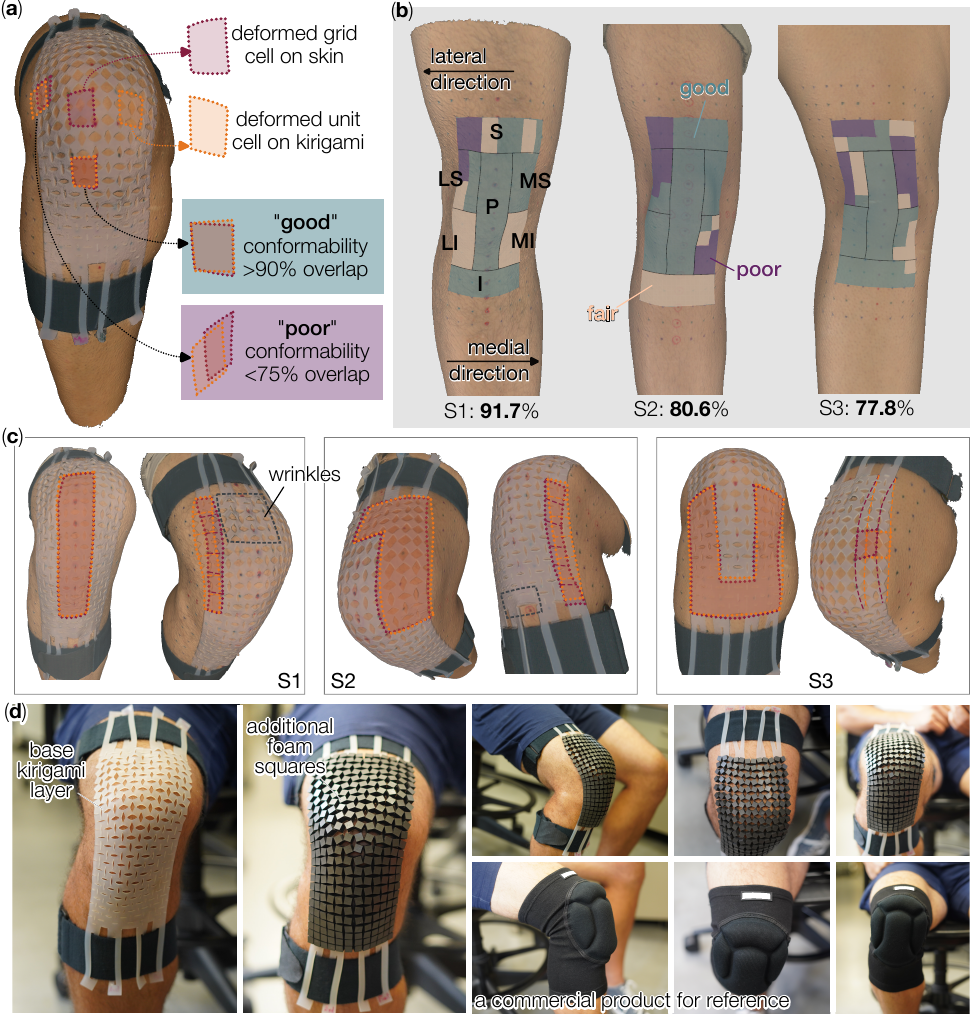}
    \caption{Conformability analysis of the three patches. (a) 3D scanning image of subject S1 wearing its personalized Kirigami patch. Here, several grid cells on the knee surface and unit cells in the Kirigami are highlighted to define conformability.  (b) A summary of conformability scores for the three Kirigami patches. (c) Close-up views highlighting a few good and poor conformability regions. (d) An application demonstration, where impact-absorbing foams are added to the silicon Kirigami to create a personalized and protective knee cap.}
    \label{fig: conform}
\end{figure}

In our design and fabrication framework, we assumed:  
1) the knee skin only stretches along the longitudinal (along the sagittal plane) and lateral (along the transverse plane) directions of the leg (Figure~\ref{fig: scan}), and  2) the unit cells in the Kirigami knee patch stretched only in-plane. 
With these two assumptions in mind, we analyzed the conformability of the final Kirigami knee patch for the three subjects. 
Here, we quantify ``conformability'' based on the areal overlap between the unit cells of the Kirigami patch over their corresponding grid cells on the subject's knee surface, when the knee is flexed at 90 degrees. 
A visual representation can be seen in Figure~\ref{fig: conform}(a), where dashed maroon lines define a grid cell drawn on the skin, and dashed orange lines define the corresponding unit cell in the deformed Kirigami patch. When the overlap between the two cells is equal to or more than 90\% of their surface area, we considered the conformability ``good.'' Meanwhile, an areal overlap greater than 75\% and less than 90\% indicates a ``fair'' conformability; and below 75\% overlap is considered ``poor.'' Figure~\ref{fig: conform} summarizes different regions of conformability on the three subjects. 

To assist in analyzing the conformability, we first divided the patch into 7 regions with respect to the patellar (Figure~\ref{fig: conform}b), including

\vspace{-8pt}
\begin{itemize}
    \item \textbf{S}: region \underline{S}uperior to the patella (top 2 rows of unit cells);
    \vspace{-8pt}
    \item \textbf{I}: region \underline{I}nferior to the patella (bottom 2 rows);
    \vspace{-8pt}
    \item \textbf{LS}: region on the \underline{L}ateral side of the leg, and \underline{S}uperior to the patella;
    \vspace{-8pt}
    \item \textbf{MS}: region on the \underline{M}edial side, and \underline{S}uperior to the patella;
    \vspace{-9pt}
    \item \textbf{LI}: region on the \underline{L}ateral side, and \underline{I}nferior to the patella;
    \vspace{-9pt}
    \item \textbf{MI}: region on the \underline{M}edial side, and \underline{I}nferior to the patella;
    \vspace{-9pt}
    \item \textbf{P} - \underline{P}atellar region, which is enclosed by the S, LS, MS, LI, MI, and I regions.
\end{itemize}
\vspace{-8pt}

With the help of these seven regions, one can more easily analyze the individual Kirigami patches and their trend of conformability performance. For example, subject S1's Kirigami patch showed some out-of-plane wrinkling in the center of region S, likely due to the concentrated stretching from the Velcro straps and fitting loops (dashed grey lines in Figure~\ref{fig: conform}c). Regardless, the overall conformability is still satisfactory among these unit cells. Some small out-of-plane wrinkling was also visible at the bottom of region I, but it did not significantly hinder this region's overall conformability either. Overall, the unit cells on the center-right portions of the Kirigami patch,  including the middle half of region S, MS, and the whole region P, showed good conformability with more than 90\% areal overlap. However, the conformability performance can start to drop towards the medial and lateral sides. For example, the outer region of LS showed an areal overlap decreasing below 75\%, as well as the lateral edge of region S.  The LI and MI regions showed a fair conformability score. Regardless, the overall conformability of the Kirigami knee patch designed for subject S1, as summarized in Figure~\ref{fig: conform}(b), still showed a promising outcome. Out of the 72 unit cells in total, 66 cells achieved at least 75\% areal overlap between the knee surface and the patch (i.e., fair and good conformability), giving a 91.67\% satisfactory ratio. 

Despite the large variation between the three subjects regarding their gender, physical activity level, and age, the performance of their Kirigami patches followed a similar trend:  Good conformability in the center with a degraded performance near the lateral and medial edges.  For subject S2,  58 cells out of 72 unit cells achieved a fair or good conformability score (an 80.56\% satisfactory rate), and for subject S3, 56 unit cells conformed satisfactorily (or 77.78\% score). These results indicate that our proposed design pipeline can accurately accommodate human body variability. 

By carefully observing the deformation mismatches between the knee skin and Kirigami in the low conformability regions (e.g., the lateral edge of region S and the upper half of region LS), we can identify several critical reasons for the discrepancy.  The first reason is the shearing deformation of the knee skin.  With a doubly curved surface that changes shape dynamically, flexing the knee joints would not only stretch the anterior knee surface, but also cause oblique distortion across the medial and lateral sides \cite{choi20153d}. However, the Kirigami cross-cut patterns adopted in this study are only designed to accommodate linear deformation along the longitudinal and lateral axes.  A new Kirigami cut template that can accommodate shearing in addition to stretching would be an interesting topic for future research. 

Another primary reason for the low conformability regions is the interaction between Kirigami unit cells with different cut designs. While simulating the Kirigami sheets' deformation and populating the design space, we assumed that all unit cells in a sheet have the same design.  However, in the final knee patch design, each cell's cut lengths are different. Therefore, there are inevitably deformation mismatches between the adjacent cells, and they might constrain each other's deformation. Moreover, the unit cells at the free boundary of the Kirigami patch or near the fitting loops could also deform differently from the FEA simulations.
These two significant discrepancy factors, in addition to the relatively minor sources of error, such as human errors involved in processing knee scanning data and the outlier targets shown in the Figure~\ref{fig: output}, all contributed to the lower conformability near the edges of the Kirigami patches. 

\subsection{An Application Case Study: Protective Knee Cap}
The conformability of the Kirigami patches allows them to deform without hindering the dynamic knee joint deformations. As a result, they can accommodate the skin deformation without peeling off or causing irritation. Moreover, the Kirigami cuts also improved the breathability of the patch by allowing perspiration to escape through the cuts.  Therefore, the Kirigami patch can be useful in many applications, such as bandages with increased adhesion (especially near moving joints), protective surfaces, or the basis for wearable electronics.

To demonstrate the practical use of conformable Kirigami knee patches, we incorporated an impact-resistant protective layer onto the silicone Kirigami and created a protective knee cap (Figure~\ref{fig: conform}d). For this purpose, we first took a 5mm, impact-resistant foam of the same size as the knee patch (Poron XRD, commonly used in sports protective gears) and cut it into squares of half the Kirigami unit cell size (i.e., 7.5mm, with Trotec Speedy 360 single point laser cutter). Notice the foam was cut completely to avoid introducing additional stiffness to the Kirigami knee patch. Then the foam sheet was carefully aligned with the unit cells of the silicon Kirigami and glued together with silicone adhesive Sil-Poxy. 
From the side view, top view, and front views in the Figure~\ref{fig: conform}(d), we can see that the protective Kirigami patch conforms well to the knee's flexion with minimal resistance. To compare with the state-of-the-art, the subject also wore a commercially available patch on the same knee (Supplemental video 2). The commercial product relied on its fabric's stretchability to support and align with the knee joint motion. This would inevitably introduce additional resistance to the body motion, and the conformability is not personalized. In contrast, the proposed protective Kirigami foam patch offers a mechanically gentler alternative, with enhanced flexibility that closely follows the skin deformation.

\section{Summary And Conclusion}
In this work, we successfully established a design and fabrication pipeline to develop a personalized body-conformable surface around the knee joint in $\sim 6.5$ hours. The process began by capturing in-plane skin deformation — from $0 \degree$ to $90 \degree$ knee flex — by measuring the longitudinal strain and Poisson's ratio of the marker grid drawn on the subjects’ anterior knee skin. Simultaneously, we used an experimentally validated FEA model of a silicone-based Kirigami sheet to train a supervised machine learning surrogate model. This training involved data from 441 FEA simulations of hyperelastic, uniformly-cut Kirigami sheets, each with a unique combination of horizontal and vertical cut lengths spanning from $1 \%$ to $99 \%$ of the unit cell size.  Then, the Gaussian process regression algorithm used these data to build a correlation between Kirigami cut lengths and the resulting longitudinal strain and Poisson's ratio accurately ($R^2 = 0.996$). Finally, an inverse design pipeline using an optimization algorithm of Covariance Matrix Adaptation Evolution Strategy (CMA-ES) and L2-normalization employed this surrogate model to generate Kirigami cut patterns that match the knee skin deformation. Implemented with parallel computing, the algorithm achieves a $92\%$ reduction in runtime. This personalized, conformable Kirigami knee patch can be laser-cut quickly from silicon rubber sheets.  We tested the fidelity of this design and fabrication pipeline on three human subjects. The Kirigami knee patches achieved areal overlapping conformability of $91.7\%, 80.6\%$, and $77.8\%$, respectively. The practical application of the Kirigami knee patch was demonstrated by integrating the personalized Kirigami sheet with a layer of impact-resistant foam, creating a conformable and protective knee cap. 

Thus, our formulated design pipeline successfully captures the complex 3D skin deformation and mitigates mechanical mismatch through a personalized, simple, low-cost, conformal skin interface. The total duration of our design and fabrication process could be further reduced by using more integrated software packages that can automatically track the grid markers on 3D knee scan data and compute their relative geodesic distances. Incorporating shear strain, alongside linear strain, and using nonlinear finite strain calculations can further improve the conformability. This design process can be readily expanded to different regions of skin, particularly areas with high curvature or joint motions. In conclusion, our proposed process enables rapid, low-cost design and fabrication of personalized and conformable surfaces that follow the complex skin deformation. They can serve as the foundation for transdermal patches, electronic skin, and wearable devices, with potential application in biosuits \cite{porter2020soft}.

\section*{Acknowledgments}
J.B. and S.L. acknowledge the partial support from the National Science Foundation (CMMI-2328522) and Virginia Tech (via startup fund and graduate student assistantship).  J.L. and J.C. acknowledge the support from Virginia Tech (via startup fund and graduate student assistantship).

\section*{Author contributions statement}
S.L. and J.C. conceived and designed this study. J.B. conducted the experiments and generated the FEA simulation data.  J.L. developed the surrogate model and design optimization pipeline.  All contributed to the drafting and revision of this manuscript.

\bibliography{references.bib}

\end{document}